\documentclass[]{JHEP3}
\usepackage{amsmath}

\renewcommand{\a}{\alpha}
\renewcommand{\b}{\beta}

\renewcommand{\d}{\delta}
\newcommand{\e}{\epsilon}
\newcommand{\ce}{\varepsilon}

\newcommand{\g}{\gamma}
\newcommand{\h}{\eta}

\renewcommand{\k}{\kappa}

\newcommand{\m}{\mu}

\newcommand{\p}{\pi}
\newcommand{\q}{\theta}

\newcommand{\s}{\sigma}

\newcommand{\w}{\omega}

\newcommand{\z}{\zeta}
\newcommand{\D}{\Delta}
\newcommand{\F}{\Phi}
\newcommand{\G}{\Gamma}

\newcommand{\ba}{\bar}
\renewcommand{\.}{\dot}
\newcommand{\pd}{\partial}
\renewcommand{\~}{\tilde}
\newcommand{\nab}{\nabla}
\newcommand{\ii}{\mathrm{i}}
\newcommand{\dd}{\mathrm{d}}

\newcommand{\no}{\nonumber}

\newcommand{\gc}[2]{\left. \lbrack #1, #2\} \right.}

\newcommand{\cd}{\mathcal{D}}

\newcommand{\1}[1][\a]{W^{#1}}
\newcommand{\2}[1][\a]{W_{#1}}
\newcommand{\3}[1][\a]{\ba{W}_{\.{#1}}}
\newcommand{\4}[1][\a]{\ba{W}^{\.{#1}}}
\newcommand{\5}{W^2}
\newcommand{\6}{\ba{W}^2}

\newcommand{\FA}{F^{ab}F_{bc}F^{cd}F_{de}F^{e}_{\phantom{e} a}}
\newcommand{\FB}{F^{ab}F_{bc}F^{cd}F^{e}_{\phantom{e} a} F_{de}}
\newcommand{\FC}{F^{ab}F^{cd}F_{bc}F^{e}_{\phantom{e} a} F_{de}}
\newcommand{\FD}{F^{ab}F^{cd}F^{e}_{\phantom{e} a} F_{bc}F_{de}}
\newcommand{\FE}{F^{ab}F_{bc}F^{c}_{\phantom{e} a} F^{de}F_{de}}
\newcommand{\FF}{F^{ab}F^{de}F_{bc}F^{c}_{\phantom{e} a} F_{de}}
\newcommand{\FGa}{(\nab^{e}F^{ab})(\nab_{e}F_{ab})F^{cd}F_{cd}}
\newcommand{\FGb}{(\nab^e F^{ab})(\nab_e F^{cd}) F_{ab} F_{cd}}
\newcommand{\FGc}{(\nab^e F^{ab})(\nab_e F^{cd}) F_{cd} F_{ab}}
\newcommand{\FMa}{(\nab^e F^{ab})(\nab_e F_{bc}) F^{cd} F_{da}}
\newcommand{\FMb}{(\nab^e F^{ab})(\nab_e F_{ca}) F_{bd} F^{dc}}
\newcommand{\FMc}{(\nab^e F^{ab})(\nab_e F^{cd}) F_{bc} F_{da}}
\newcommand{\FMd}{(\nab^e F^{ab}) F_{da}(\nab_e F_{bc}) F^{cd}}
\newcommand{\FGd}{(\nab^e F^{ab}) F^{cd}(\nab_e F_{ab}) F_{cd}}
\newcommand{\FLa}{(\nab^e F^{ab}) F^{cd}(\nab_a F_{bc}) F_{de}}
\newcommand{\FLb}{F^{ab}(\nab_a F^{bc}) F^{de}(\nab_e F_{cd})}
\newcommand{\FLc}{F^{ab}(\nab_a F^{cd})(\nab^e F_{bc}) F_{de}}

\newcommand{\perm}[4]{ #1{\1}\Big( #2{\2} #3{\3} #4{\4}+ #2{\3}
#3{\4} #4{\2}- #2{\3} #3{\2} #4{\4} \Big)}
\newcommand{\cda}[2]{(#1{#2})}

\title{Higher order contributions to the effective action
of $\mathcal{N}=4$ super Yang-Mills}

\author{Darren T. Grasso\\ School of Physics,
The University of Western Australia \\Crawley, W.A. 6009.
Australia\\E-mail: \email{grasso@physics.uwa.edu.au}}

\abstract{The one-loop low-energy effective action for non-Abelian
$\mathcal{N}=4$ supersymmetric Yang-Mills theory is computed to
order $F^6$ by use of  heat kernel techniques in $\mathcal{N}=1$
superspace. At the component level, the $F^5$ terms are found to
be consistent with the form of the non-Abelian Born-Infeld action
computed to this order by superstring methods. The $F^6$ terms
will be of importance for comparison with superstring
calculations.}

\begin{document}

\section{Introduction}
There is considerable interest in the issue of deformations of
maximally supersymmetric Yang-Mills theories
\cite{BRS,MR,MRT,T0,PSS1,CNT1}; for a review see \cite{T1}. Such
deformations arise in the context of superstring theories, where
the low-energy effective actions for D-branes admit expansions in
powers of the string tension\footnote{The generic structure is of
the form $\sum_{n=0}^{\infty} c_n \,(\a')^{n} \, F^{n+2}$, where
$F^{n}$ denotes  terms of mass dimension $2n$ in $F$ and its
covariant derivatives. For D-brane probes in the background of a
stack of D-branes, the expansion parameter is not $\a ',$ but is
determined by the vacuum expectation values of scalar fields which
specify the separation of the probe from the stack.}  $\a '$. The
lowest order term is a maximally supersymmetric Yang-Mills action.
For a single D-brane, the terms in the low-energy effective action
which do not contain derivatives of the field strength are known
to all orders in $\a '$: they are given by the Born-Infeld action
\cite{FT,L}. In the case when there are N coincident D-branes, the
resulting low-energy effective action has been dubbed the
non-Abelian Born-Infeld action \cite{T2}, because the lowest order
term in the expansion is the action for SU(N) supersymmetric
Yang-Mills theory \cite{W1}, and it reduces to the Born-Infeld
action in the Abelian case. Due to the Bianchi identity
\begin{equation}\label{eq:bianchi}
[F_{a b},F_{c d}] = 2 \ii \, \nab_{[a} \nab_{b]}F_{c d},
\end{equation}
it is not possible to consistently truncate the non-Abelian
Born-Infeld action to constant field strength, and so derivative
corrections must be considered \cite{T1}. As yet, only a few terms
in the $\a'$ expansion of the non-Abelian Born-Infeld action are
known.

Of particular interest is the question as to whether supersymmetry
is a sufficiently strong constraint to uniquely specify  the form
(up to field redefinitions) of the deformation of a maximally
supersymmetric Yang-Mills theory \cite{T0,PSS1}. If this were the
case, then any means to compute a supersymmetric deformation would
yield the non-Abelian Born-Infeld action. In particular, low
energy effective actions for supersymmetric Yang-Mills theories
would be related to the non-Abelian Born-Infeld action. However,
since the effective action is dependent on the choice of gauge,
with a change of gauge inducing a field redefinition, direct
comparison of low-energy effective actions with deformations
obtained by other means is potentially nontrivial.

The $F^5$ contributions to the non-Abelian Born-Infeld action have
recently been calculated in full by several different methods
\cite{refolli, KS1,MBM}. In \cite{refolli}, the one-loop
low-energy effective action for $\mathcal{N}=4$ supersymmetric
Yang-Mills theory in four dimensions was calculated using
supergraphs, and the result was:
\begin{multline}\label{eq:refolliresult}
\k_1 \,\mathrm{Tr}\bigg(2\FMa+2\FMc+\FMb\\
-\frac{1}{2}\FGa-\frac{1}{2}\FGb-\frac{1}{2}\FGc \\
+2\ii \FA +7\ii \FB+ 3\ii \FD -4\ii \FC\bigg),
\end{multline}
where $\k_1$ is a normalization. On the other hand, Koerber and
Sevrin \cite{KS1} used an approach based on the requirement that
certain BPS solutions should exist to the equations of motion
derived from the non-Abelian D-brane effective action, extending
earlier use of this method for the Abelian case \cite{FKS}. At
order $(\alpha')^3$, this approach yields\footnote{Partial results
at order $(\a')^3$ had previously been obtained in
\cite{tseytlin}, \cite{B} and \cite{K}.  $F^5$ terms in ten
dimensional super Yang-Mills were given in \cite{vdv}.}
\begin{gather}
\k_2\,\mathrm{Tr}\bigg(\FLc-\FLa-\frac{1}{2}\FMb
\no \\ \label{eq:NABIresult} -\frac{1}{2}\FMd +\frac{1}{8}\FGd \\
-\frac{\ii}{10}\FA-\frac{\ii}{2}\FB
-\frac{\ii}{2}\FC+\frac{\ii}{7}\FD \bigg), \no
\end{gather}
where $\k_2$ is a normalization constant.  Comparison of
(\ref{eq:refolliresult}) and (\ref{eq:NABIresult}) is not
straightforward due to the identity (\ref{eq:bianchi}). However,
when written using the basis for the various tensor structures
adopted in (\ref{eq:NABIresult}), the result
(\ref{eq:refolliresult}) takes the form
\begin{gather}
8\k_1\,\mathrm{Tr}\bigg(\FLc-\FLa-\frac{1}{2}\FMb
\no \\ \label{eq:refolliresult2} -\frac{1}{2}\FMd +\frac{1}{8}\FGd \\
+\frac{3 \ii}{20}\FA+\frac{\ii}{8}\FB
-\frac{\ii}{2}\FC+\frac{33\ii}{40}\FD \bigg), \no
\end{gather}
As can be seen, the terms  containing covariant derivatives of the
field strength coincide with (\ref{eq:NABIresult}), but the terms
without covariant derivatives differ \cite{KS1,KS2}.

A number of tests have successfully been applied to confirm that
expression (\ref{eq:NABIresult}) is consistent with string
theoretic predictions \cite{KS2,REKS}. Most recently, a string
theory calculation of the full five-point scattering amplitude for
gluons has been carried out  \cite{MBM}, from which it is inferred
that the corresponding low-energy effective action has precisely
the order $(\a ')^3$ terms (\ref{eq:NABIresult}). This technique
was first applied  at order $(\a')^2$ by Gross and Witten
\cite{GW}. Other approaches have also provided information on the
Born-Infeld action at this order \cite{cederwall,deroo,STT}.

In this paper, the one-loop low-energy effective action for
non-Abelian $\mathcal{N}=4$ supersymmetric Yang-Mills theory is
calculated in $\mathcal{N}=1$ superfield form through to order
$F^6.$ The technique employed is a modification of that developed
in \cite{mcarthur, gargett} based on the properties of `moments'
of heat kernels\footnote{For an alternative technique, see
\cite{pletnev,banin}.}. At order $F^5,$ extraction of components
from the resulting superfield expression yields
(\ref{eq:NABIresult}), rather than (\ref{eq:refolliresult}). The
fact that three different means to calculate a $F^5$ deformation
of supersymmetric Yang-Mills theory yield the same result
(\ref{eq:NABIresult}) is evidence for  the existence of a unique
deformation at this order. If indeed this is the case, it also
suggests that the $F^5$ terms in the low-energy effective action
for $\mathcal{N}=4$ supersymmetric Yang-Mills theory are not
renormalized beyond one-loop. The one-loop non-Abelian $F^6$ terms
computed in this paper are potentially important for comparison
with recent string theoretic results \cite{koerber,ST1,ST2}, as
are the recently computed two-loop Abelian $F^6$ terms \cite{BPT}.

The paper is organized as follows.  Firstly the quantization of
non-Abelian $\mathcal{N}=4$ supersymmetric Yang-Mills theory in
$\mathcal{N}=1$ superfield form is briefly reviewed, including the
background field method, heat kernels and zeta function
regularization.  In Section \ref{sec:DE} a general expression
adapted to the asymptotic expansion of the heat kernel is derived.
This expression is then used in Section \ref{sec:subleading} to
compute the $F^5$ terms in the low-energy effective action in
superfield form, from which we extract the bosonic component and
make comparisons with the results (\ref{eq:refolliresult}) and
(\ref{eq:NABIresult}).  In Section \ref{sec:a6} we explain how the
technique generalizes to allow a computation of the $F^6$ terms in
the one-loop low-energy effective action, and the full superfield
result can be found in the Appendix B. Some comments on the form
of the result are included in Section \ref{sec:discussion}.

We adopt the conventions and notation of \cite{Wess} and \cite{kuzenko}.

\section{Quantization of $\mathcal{N}=4$ super Yang-Mills
\label{sec:quantization}}

The non-Abelian $\mathcal{N}=4$ supersymmetric Yang-Mills action
cast in $\mathcal{N}=1$ superfield form is
\begin{eqnarray}
S=\frac{1}{g^2}\mathrm{Tr}\bigg(\int\!\!\dd^{8}z \;
e^{-2V}\ba{\F}_i e^{2V}\F_i+\frac{1}{4}\int\!\!\dd^{6}z\;
\1\2+\frac{1}{4}\int\!\!\dd^{6}\ba{z}\;\3\4 \no \\
+\frac{\sqrt{2}}{3!}\int\!\!\dd^{6}z\;\e^{ijk}[\F_i,\F_j] \F_k
+\frac{\sqrt{2}}{3!}\int\!\!\dd^{
6}\ba{z}\;\e^{ijk}[\ba\F_i,\ba\F_j] \ba\F_k\bigg),
\end{eqnarray}
where
\begin{equation}\2=-\frac{1}{8}\ba{D}^2(e^{-2V}D_{\a}
e^{2V})
\end{equation}
and $\F_i$ ($i=1,\,2,\,3$) are chiral superfields.  All
superfields are Lie-algebra valued, for example
$\F_i=\F_i^I\,T^I$, with the Hermitian generators $T^I$
satisfying:
\begin{equation}
[T^I,\,T^J]=\ii f^{IJK}\,T^K.
\end{equation}

In the $\mathcal{N}=1$ background field formalism for
supersymmetric non-Abelian gauge theories \cite{grisaru}, it is
necessary to introduce a nonlinear background-quantum splitting to
ensure a gauge invariant effective action. Defining
\begin{equation}
e^{2V} \equiv e^{w} e^{\ba{w}},
\end{equation}
the background-quantum splitting is given by
\begin{equation}
e^{w}=e^{w_B}e^{{w}_Q},
\end{equation}
or equivalently
\begin{equation}
e^{2V}=e^{w_B}e^{2V_Q}e^{\ba{w}_B}
\end{equation}
where the subscripts $B$ and $Q$ denote background and quantum
pieces respectively. Since we will be performing a one-loop
calculation it is only necessary to retain terms quadratic in
quantum fields in the action.  After background covariant gauge
fixing \cite{grisaru,grisaru2}, the quantum quadratic action for
$\mathcal{N}=4$ super Yang-Mills becomes
\begin{equation}\label{eq:quadraticaction}
S_{quad}=\int\!\! \dd^{8}z\; V_{Q} \; (\D_1-m^2)\; V_{Q}
\end{equation}
where
\begin{equation}
\D_1= \cd^{a} \cd_{a}-W_B^{\a}\cd_\a
-\ba{W}_B^{\.{\a}}\ba{\cd}_{\.{\a}}.
\end{equation}
The mass $m^2=|\F_B|^2$ is introduced via a constant chiral scalar
superfield background $\F_B,$ and the low-energy effective action
has an expansion in inverse powers of $m^2.$ The $W_B$'s and
$\cd$'s are background superfield strengths and background gauge
covariant derivatives respectively, defined and related by
(dropping all $B$ subscripts):
\begin{gather}
\cd_\a =  e^{-w}D_{\a}e^{w}, \qquad \ba{\cd}_{\.{\a}} =
e^{\ba{w}}\ba{D}_{\.\a}e^{-\ba{w}} \no \\
\cd_a = -\frac{1}{2}({\~{\s}}_a)^{\.{\a} \a}\cd_{\a
\.{\a}}=-\frac{\ii}{4}({\~{\s}}_a)^{\.{\a}
\a}\{\cd_\a,\ba{\cd}_{\.{\a}}\} \\
[\ba{\cd}_{\.{\a}},\cd_{\b \.{\b}}]=2 \ii \ce_{\.{\a}
\.{\b}}W_{\b}, \qquad [\cd_{\a},\cd_{\b {\.{\b}}}]=2 \ii \ce_{\a
\b}\ba{W}_{\.{\b}}. \no
\end{gather}

One can efficiently calculate the one-loop effective action
through zeta function regularization, and it is just
$-\frac{1}{2}\,\z'(0)$, where the zeta function is defined by
\begin{equation}\label{eq:zeta}
\z(s)=\frac{\m^{2s}}{\G(s)}\int_0^\infty\!\!\dd t \; t^{s-1} e^{-t
m^2} K(t).
\end{equation}
In this expression $\m$ is the renormalization point and $K(t)$ is
the functional trace of the heat kernel associated with the
operator $\D_1$,
\begin{equation}
K(t)= \mathrm{Tr}\int\!\! \dd^{8}z \,\lim_{z'\rightarrow z} e^{t
\D_1} \d^{(8)} (z,z') \equiv \mathrm{Tr}\int\!\! \dd^{8}z
\,\lim_{z'\rightarrow z} K(z,z',t).
\end{equation}
Here $\mathrm{Tr}$ denotes the trace over gauge indices and
$\d^{(8)} (z,z')$ is the superspace delta function,
\begin{equation}
\d^{(8)}(z,z')=\d^{(4)}(x,x')\d^{(2)}(\q-\q')\d^{(2)}(\ba{\q}-\ba{\q}').
\end{equation}
Introducing a plane wave basis for the delta functions,
\begin{gather}
\d^{(4)}(x,x')=\int\!\!\frac{\dd^{4}k}{(2 \p)^4}\; e^{\ii k^a
\w_a}\\ \d^{(2)}(\q-\q')= 4\int\!\!\dd^{2} \e \;e^{\ii
\e^\a(\q-\q')_\a}, \qquad \d^{(2)}(\ba{\q}-\ba{\q}')=
4\int\!\!\dd^{2}\ba{\e} \; e^{\ii
\ba{\e}_{\.{\a}}(\ba{\q}-\ba{\q}')^{\.{\a}}}
\end{gather}
where
\begin{equation}
\w_a=x_a-x'_{a}-i \q \s_a \ba{\q}'+i \q' \s_a \ba{\q},
\end{equation}
and defining
\begin{equation}\label{eq:measure}
\int\!\!\dd \h = 16 \int\!\!\frac{\dd^{4}k}{(2
\p)^4}\int\!\!\dd^{2}\e \int\!\!\dd{^2}\ba{\e},
\end{equation}
$K(z,z',t)$ becomes
\begin{equation}\label{eq:originalkernel}
K(z,z',t)=\int\!\!\dd \h \;e^{\ii k^a \w_a} e^{\ii
\e^\a(\q-\q')_\a} e^{\ii
\ba{\e}_{\.{\a}}(\ba{\q}-\ba{\q}')^{\.{\a}}} e^{ t \D}
\end{equation}
with
\begin{equation}
\D=X^a X_a-W^{\a}X_{\a}-\ba{W}^{\.{\a}}\ba{X}_{\.{\a}}.
\end{equation}
Here, the $X$'s are defined by
\begin{align}
X_a&=\cd_a+i k_a \no \\
X_{\a}&=\cd_{\a}+i\e_\a-k_{\a\.{\a}}(\ba{\q}-\ba{\q}')^{\.{\a}}
\\ \ba{X}_{\.{\a}}&=\ba{\cd}_{\.{\a}}+i\ba{\e}_{\.{\a}}
+k_{\a\.{\a}}(\q-\q')^{\a},\no
\end{align}
and satisfy the algebra
\begin{gather}
\{X_\a,X_\b\}=\{\ba{X}_{\.{\a}},\ba{X}_{\.{\b}}\}=0, \quad \no
\{X_\a,\ba{X}_{\.{\a}}\}=-2iX_{\a \.{\a}}, \quad [X_a,X_b]=G_{ab}
\no\\ [X_\a,X_{\b\.{\b}}]=2i \ce_{\a \b}\ba{W}_{\.{\b}}, \qquad
[\ba{X}_{\.{\a}},X_{\b\.{\b}}]=2i \ce_{\.{\a}\.{
\b}}W_{\b}\label{eq:commutation}
\\ G_{\a\.{\a},\b\.{\b}}=(\s^a)_{\a\.{\a}}(\s^b)_{\b\.{\b}}G_{ab}
=-\ce_{\a\b}(\ba{\cd}_{\.\a}\ba{W}_{\.\b})-\ce_{\.{\a}\.{\b}}(\cd_\a{W}_\b).\no
\end{gather}
Taking the limit\footnote{This limit is implicitly taken
throughout the remainder of this paper.} $z'\rightarrow z$ in
(\ref{eq:originalkernel}), one obtains
\begin{equation}\label{eq:kerneldefinition}
K(z,t)\equiv \lim_{z'\rightarrow z}K(z,z',t)=\int\!\!\dd \h \;e^{
t \D}.
\end{equation}

The kernel $K(z,t)$ has an asymptotic expansion in $t$ in the
limit $t\rightarrow 0$, and the leading term in the expansion is
of order $t^2$.  This fact can be seen by making the rescaling
$k_a\rightarrow t^{-\frac{1}{2}}k_a$, and by observing that the
integral over the fermionic parameters $\e_\a$ and
$\ba{\e}_{\.{\a}}$ will bring down at least four factors of $t$.
Defining the DeWitt-Seeley coefficients $a_n$ in the usual manner,
\begin{equation}\label{eq:asymptotic}
K(z,t)=\frac{1}{16 \p^2 t^2}\sum_{n=0}^{\infty}t^{n} a_n  \qquad
a_i = 0, \quad i=0,1,2,3,
\end{equation}
the one-loop effective action then takes the form
\begin{equation}\label{eq:effectiveaction}
\G_{(1)} \equiv - \frac{1}{2} \, \z'(0)=-\frac{1}{32 \p^2}
\sum_{n=4}^\infty \frac{(n-3)!}{m^{2n-4}}\int\!\! \dd^{8}z \,
\mathrm{Tr}(a_n),
\end{equation}
which is an expansion in inverse powers of the mass parameter.  At
the component level, the non-trivial DeWitt-Seeley coefficients,
$a_n$ for $n\geq4$, contain bosonic field strength terms of the
form $F^n$. The first non-trivial coefficient, $a_4,$ is
well-known (see for example \cite{GGRS,ohrndorf}):
\begin{equation}
\mathrm{Tr}(a_4) = \frac{1}{3}\, \mathrm{Tr}(2\, \5 \6- \1 \3 \2
\4).
\end{equation}
Our goal is to compute $a_5$ and $a_6$ in superfield form.

In general the process of asymptotically expanding heat kernels is
involved and very laborious. The most direct route, expanding the
exponential $e^{t\D}$, is cumbersome and really only practical for
computing the first non-trivial DeWitt-Seeley coefficient. To
calculate higher order coefficients it is necessary to introduce
an efficient algorithm, which is done in the next section.

\section{The differential equation approach \label{sec:DE}}
We proceed by modifying the differential equation approach
developed in \cite{mcarthur,gargett}.  Briefly, this approach
involves generating a differential equation for $K(z,t)$.  By
exploiting certain properties of the kernel, and provided the
background is sufficiently simple, the resulting equation can be
solved either iteratively or in some cases exactly by expressing
$\frac{d K(z,t)}{d t}$ in terms of $K(z,t)$. For more complicated
backgrounds, as with the case at hand, it becomes too difficult to
express the differential equation in a form in which it can be
solved.

However, the techniques employed in \cite{mcarthur,gargett}  may
be used in a rather different way to facilitate the computation of
higher order DeWitt-Seeley coefficients in theories with arbitrary
backgrounds. In this new approach, one does not actually attempt
to solve the differential equation, as we now illustrate.

We begin by differentiating $K(z,t)$ with respect to $t$, which
yields the differential equation
\begin{equation}\label{eq:de1}
\frac{d K(z,t)}{d t}=
K^a_{\phantom{a}a}(z,t)-W^{\a}K_{\a}(z,t)-\ba{W}^{\.{\a}}K_{\.{\a}}(z,t),
\end{equation}
where the notation
\begin{equation}\label{eq:kernel}
K_{A_1 A_2 \ldots A_n}(z,t)=\int\!\! \dd \h \;
X_{A_1}X_{A_2}\ldots X_{A_n} e^{t\D}
\end{equation}
has been introduced, with the integration measure defined in
(\ref{eq:measure}). Using the identities
\begin{equation}\label{eq:id1}
0=\int\!\! \dd \h\; \frac{\pd \phantom{k_b}}{\pd k_b}\left(X_a
e^{t\D}\right)
\end{equation}
and
\begin{equation}
[A, e^B]=\int_{0}^{1}\!\!\dd s\;e^{s B}[A,B]e^{(1-s)B},
\end{equation}
it follows that
\begin{equation}
0=\ii \d_a^b K(z,t)+2 \ii t\int\!\!\dd
\h\;X_a\sum_{n=0}^{\infty}\frac{t^n}{(n+1)!} ad_{\D}^{\, n}(J^b)
\;e^{t\D}
\end{equation}
where
\begin{equation}
J^a=X^a-\frac{\ii}{2}W \s^a
(\ba{\q}-\ba{\q}')-\frac{\ii}{2}(\q-\q')\s^a \ba{W}
\end{equation}
and $ad^{n}$ denotes $n$ nested commutators:
\begin{equation}
ad_{A}^{\, 0}(B)=B, \qquad
ad_{A}^{\,n}(B)=[A,\,ad_{A}^{\,n-1}(B)].
\end{equation}
After contraction of vector indices, this becomes
\begin{equation}\label{eq:k1}
K^a_{\phantom{a}a}(z,t)=-\frac{2}{t}K(z,t) -\int\!\!\dd
\h\;X_a\sum_{n=1}^{\infty}\frac{t^n}{(n+1)!}
ad_{\D}^{\,n}(J^a)\;e^{t\D}.
\end{equation}

Similarly, using
\begin{equation}\label{eq:id2}
0=\int\!\! \dd \h \; \frac{\pd \phantom{\e_{\b}}}{\pd
\e_\b}\left(X_{\a} e^{t\D}\right)
\end{equation}
and
\begin{equation}\label{eq:id3} 0=\int\!\! \dd \h \; \frac{\pd
\phantom{\ba{\e}_{\.{\b}}}}{\pd\ba{\e}_{\.{\b}}}\left(\ba{X}_{\.{\a}}
e^{t\D}\right)
\end{equation}
it follows that
\begin{equation}\label{eq:k2}
W^{\a}K_{\a}(z,t)=-\frac{2}{t} K(z,t)+(\cd_{\a}W^\a)K(z,t)
+\int\!\!\dd \h\;X_{\a}\sum_{n=1}^{\infty}\frac{t^n}{(n+1)!}
ad_{\D}^{\,n}(W^\a)\;e^{ t\D}
\end{equation}
and
\begin{equation}\label{eq:k3}
\ba{W}^{\.{\a}} K_{\.{\a}}(z,t)=-\frac{2}{t}K(z,t)+
(\ba{\cd}_{\.{\a}} \ba{W}^{\.{\a}})K(z,t) +\int\!\!\dd
\h\;\ba{X}_{\.{\a}}\sum_{n=1}^{\infty}\frac{t^n}{(n+1)!}
ad_{\D}^{\,n}(\ba{W}^{\.{\a}})\;e^{t\D}
\end{equation}
respectively.

Finally, inserting (\ref{eq:k1}),(\ref{eq:k2}) and (\ref{eq:k3})
into the differential equation (\ref{eq:de1}) one obtains:
\begin{align}
\frac{d K(z,t)}{d t}-\frac{2}{t}K(z,t)=&-
\int\!\!d\h\;X_a\sum_{n=1}^{\infty}\frac{t^n}{(n+1)!}
ad_{\D}^{\,n}(J^a)\;e^{t\D}\no
\\&-\int\!\!d\h\;X_{\a}\sum_{n=1}^{\infty}\frac{t^n}{(n+1)!}
ad_{\D}^{\,n}(W^\a)\;e^{t\D} \no
\\&-\int\!\!d\h\;\ba{X}_{\.{\a}}\sum_{n=1}^{\infty}
\frac{t^n}{(n+1)!}ad_{\D}^{\,n}(\ba{W}^{\.{\a}})\;e^{t\D},\label{eq:de2}
\end{align}
where the $K(z,t)$ pieces have been brought to the left hand side
and the Bianchi identity,
$\cd^{\a}W_\a=\ba{\cd}_{\.{\a}}\ba{W}^{\.{\a}}$, has been used.
The significance of this expression is seen in terms of the
asymptotic expansion (\ref{eq:asymptotic}), where the left hand
side is
\begin{equation}\label{eq:powerseries}
\frac{d K(z,t)}{d t}-\frac{2}{t}K(z,t)=\frac{1}{16
\p^2}\sum_{n=0}^{\infty}\, (n-4) t^{n-3} a_n=\frac{a_5 t^2}{16
\p^2}+\frac{2 a_6t^3}{16 \p^2}+\cdots
\end{equation}
It is clear that in this particular combination of the kernel and
its derivative, the first non-trivial coefficient, $a_4$, is
absent\footnote{This feature is not particular to the current
example.  Differential equations of the form (\ref{eq:de2}), where
the first non-trivial coefficient is absent, arise naturally when
applying these techniques to heat kernels associated with
`reasonable' operators in superspace with arbitrary dimensions.
This is most obvious in ordinary p-dimensional spacetime with
Laplace-type operators.}. Exploiting this fact, the objective now
becomes to determine the DeWitt-Seeley coefficients by expanding
the right hand side of (\ref{eq:de2}) in a power series in $t$,
and identifying it with the right hand side of
(\ref{eq:powerseries}).

The background has been arbitrary to this point. However, from now
on it will be placed on-shell,
$\cd^{\a}W_\a=\ba{\cd}_{\.{\a}}\ba{W}^{\.{\a}}= 0$, as we are only
interested in the on-shell effective action. Since the summation
on the right hand side of (\ref{eq:de2}) involves the repetitive
calculation of commutators, it is first useful to establish the
following relations:
\begin{eqnarray}
\lbrack\D,X_a\rbrack &=& 2 G^b_{\phantom{b}a}X_b + (\cd_a
W^\a)X_\a+(\cd_a \ba{W}^{\.{\a}})\ba{X}_{\.{\a}} \no \\ \lbrack
\D,X_\a \rbrack &=&(\cd_\a W^\b)X_\b \no \\ \lbrack
\D,\ba{X}_{\.{\a}}\rbrack &=& (\ba{\cd}_{\.{\a}}
\ba{W}^{\.{\b}})\ba{X}_{\.{\b}}\label{eq:commutation2} \\ \lbrack
\D,A\rbrack &=&(\cd^a\cd_a A)+2(\cd^a A)X_a-W^\a(\cd_\a
A)-\ba{W}^{\.{\a}}(\ba{\cd}_{\.{\a}}A) \no \\&&-(-1)^{\ce(A)}
\gc{W^\a}{A}X_\a-(-1)^{\ce(A)}\gc{\ba{W}^{\.{\a}}}{A}\ba{X}_{\.{\a}}.\no
\end{eqnarray}
From these it is clear that summation will generate a series of
objects of the form $K_{A_{1} \ldots A_{i}}(z,t)$, which we shall
refer to as moments of the kernel\footnote{Occasionally we also
use this term to collectively refer to all moments and $K(z,t)$
itself.}, as defined in (\ref{eq:kernel}). Furthermore, it is not
difficult to show that to order $n$ in this summation, the moments
generated have at most $(n+1)$ indices. It is convenient to always
place these indices in a specific order: first undotted, then
dotted, then spacetime. This can be achieved through the
commutation relations (\ref{eq:commutation}). With such an
ordering, the leading term in a moment's asymptotic power series
has the following behaviour\footnote{With arbitrary ordering the
behaviour is only slightly more complicated.}:
\begin{equation}
K_{A_1 \ldots A_{p+q}}(z,t)\sim \frac{1}{t^2}
\left(\frac{1}{t}\right)^{\left[\frac{p}{2}\right]}t^{4-q} =
t^{2-q-\left[\frac{p}{2}\right]}\qquad \qquad q\leq4
\end{equation}
where $K_{A_1 \ldots A_{p+q}}(z,t)$ has $p$ spacetime indices, $q$
spinor indices and $[\frac{p}{2}]$ denotes the largest integer
part of $\frac{p}{2}$. Moments with greater than two undotted or
dotted indices vanish as $X_\a X_\b X_\g=\ba{X}_{\.{\a}}
\ba{X}_{\.{\b}} \ba{X}_{\.{\g}}=0$.

From these considerations, and by comparison with equation
(\ref{eq:powerseries}), the summation in equation (\ref{eq:de2})
truncates at $n=2k-5$ when evaluating $a_k$ for $k\geq5$.
Moreover, it turns out that after tracing over gauge indices, the
last term in this truncated summation always vanishes due to the
cyclic property of the trace, making it necessary to sum only to
$n=2k-6$. In particular, this means that to evaluate $a_5$ and
$a_6$ one is permitted to truncate at $n=4$ and 6 respectively.
More explicitly, the terms with $n=2k-5$ are always of the form
\begin{equation}
t^{2k-5}M^{\a\b\.{\a}\.{\b}b_1 \ldots b_{2k-8}}
K_{\a\b\.{\a}\.{\b}b_1 \ldots b_{2k-8}}(z,t)\qquad \qquad k\geq5
\end{equation}
where the coefficient $M$ is some graded commutator.  The moment
in this expression is only ever required to leading order in $t,$
and at this order, it is proportional to the identity matrix in
its group indices. Consequently all contributing terms are only
proportional to graded commutators, which vanish under the trace.

\section{The $F^5$ terms \label{sec:subleading}}
In this section, the calculation of the DeWitt-Seeley coefficient
$a_5$ in $\mathcal{N}=4$ supersymmetric Yang-Mills theory is
discussed. This determines the $F^5$ terms in the low-energy
effective action.

\subsection{Evaluating $a_5$}
Before proceeding, it is instructive to examine the differential
equation (\ref{eq:de2}) in a little more detail. Since
\begin{equation}
ad_{A}^{\,n}(BC)=\sum_{m=0}^{n}\frac{n!}{m!(n-m)!}\;
ad_{A}^{\,n-m}(B)\; ad_{A}^{\,m}(C),
\end{equation}
if $A$ has even Grassmann parity, then
\begin{align}
ad_{\D}^{n}(J^a-X^a)&=\frac{\ii}{2}(\s^a)_{\a \.{\a}}
ad_{\D}^{\,n}((\ba{\q}-\ba{\q}')^{\.{\a}}W^\a-(\q-\q')^\a
\ba{W}^{\.{\a}}) \no \\ &=\frac{\ii}{2}(\s^a)_{\a \.{\a}}
\big((\ba{\q}-\ba{\q}')^{\.{\a}}ad_{\D}^{\,n}(W^\a)-(\q-\q')^\a
ad_{\D}^{\,n}(\ba{W}^{\.{\a}})\big)  + M_n
\end{align}
where
\begin{equation*}
M_n=\frac{\ii}{2}(\s^a)_{\a \.{\a}}\sum_{m=0}^{n-1}
\frac{n!}{m!(n-m)!} \big(
ad_{\D}^{\,n-m-1}(\ba{W}^{\.{\a}})\;ad_{\D}^{\,m}(W^\a)+
ad_{\D}^{\,n-m-1}(W^\a)\;ad_{\D}^{\,m}(\ba{W}^{\.{\a}})\big).
\end{equation*}
The differential equation (\ref{eq:de2}) can then be expressed in
the more useful form
\begin{align}
\frac{d K(z,t)}{dt}-\frac{2}{t}K(z,t)=&-\int\!\!\dd
\h\;X_a\sum_{n=1}^{\infty}\frac{t^n}{(n
+1)!}ad_{\D}^{\,n}(X^a)\;e^{t\D} \no\\&-\int\!\!\dd
\h\;X_{\a}\sum_{n=1}^{\infty}\frac{t^n}{(n+1)!}ad_{\D}^{\,n}
(W^\a)\;e^{t\D}\no\\&-\int\!\!\dd
\h\;\ba{X}_{\.{\a}}\sum_{n=1}^{\infty}\frac{t^n}{(n+1)!}ad_{\D}^{\,n}
(\ba{W}^{\.{\a}})\;e^{t\D} \no\\&-\int\!\!\dd
\h\;X_a\sum_{n=1}^{\infty}\frac{t^n}{(n+1)!}M_n \;e^{t\D}.
\label{eq:de3}
\end{align}

To order $n$ in the summation, the last of the four terms on the
right hand side will generate moments with at most $n$ indices
(whereas the first three generate moments with at most $n+1$). By
investigating its powers series behaviour, one ultimately finds
that this last term will not contribute when computing $a_5$, and
can safely be ignored.

To illustrate the manner in which the first three terms are
evaluated, consider the $n=1$ contribution to the second term. Use
of the commutation relations (\ref{eq:commutation2}) gives:
\begin{align}
-\frac{t}{2!}\int\!\!\dd
\h\;X_{\a}[\D,W^\a]\;e^{t\D}=&-\frac{t}{2!}\Big((\cd_\a
A^\a)K(z,t)+((\cd_\b C^{\b \a})-A^\a)K_\a (z,t) \no \\& +(\cd_\a
E^{\a\.{\a}})K_{\.{\a}}(z,t) +(\cd_\a B^{\a a})K_a (z,t) \no \\&
-B^{\a a}K_{\a a}(z,t)+C^{\a \b}K_{\a \b}(z,t)+E^{\a \.{\a}} K_{\a
\.{\a}}(z,t) \Big)
\end{align}
with
\begin{equation}
\begin{split}
A^\a =(\cd^a \cd_a W^\a)-W^\b (\cd_\b W^\a), \qquad B^{\a
a}=2(\cd^a W^\a) \\C^{\a \b}=\{W^\a,W^\b\}, \qquad E^{\a
\.{\a}}=\{W^\a,\ba{W}^{\.{\a}}\}.\qquad
\end{split}
\end{equation}
This is further simplified by the vanishing of the
coefficient of $K_{\.{\a}}(z,t)$ due to chirality and the
equations of motion, and $C^{\a \b}K_{\a \b}(z,t)$ vanishes due to
the symmetry/antisymmetry of its indices.  Furthermore, the terms
involving $t\,K(z,t)$ and $t\,K_a(z,t)$ will not contribute to the
order of interest, since after expansion both have will have
leading terms of order $t^3$. Curiously, and apparently contrary
to equation (\ref{eq:powerseries}), one also finds a term, $t\,
K_{\a \.{\a}}(z,t)$, of leading order $t$, but all such terms are
found to cancel (as they must) when considering the complete right
hand side of (\ref{eq:de3}).

There now remains the problem of expanding the contributing
moments to the required order in $t$.  In the current example this
does not pose any additional difficulties since it is only
necessary to expand all surviving moments to leading order, and as
explained above, such expressions are readily obtained by directly
expanding the exponential.  For example:
\begin{equation}
K_{\a}(z,t) = \int\!\! \dd \h \; X_{\a} e^{t\D}=-\frac{1}{16 \p^2}
\frac{4 t}{3!}(W_\a \ba{W}^2+ \ba{W}^2
W_\a-\ba{W}_{\.{\a}}W_\a\ba{W}^{\.{\a}})+\mathcal{O}(t^2)
\end{equation}
where the $k$ integral has also been performed (after Wick
rotation to a Euclidean metric). In essence the problem of
computing the kernel to subleading order has been reduced to
computing several moments to leading order.

Carrying out this procedure to order $t^2$ for the entire right
hand side of equation (\ref{eq:de3}) for $n=1$ to 4, and using the
cyclicity of the trace, $\mathrm{Tr}(a_5)$ can be identified:
\begin{multline}\label{eq:2nd}
\mathrm{Tr}(a_5)= \frac{1}{90}\mathrm{Tr}\bigg(10 \Big( (\cd^a
\cd_a \1)\2 \6+ (\cd^a \cd_a \1)\6 \2 - (\cd^a \cd_a \1)\3 \2 \4
\Big)\\ +11 \Big( (\cd^a \1)(\cd_a \2) \6+ (\cd^a \1)(\cd_a \3) \4
\2 - (\cd^a \1)(\cd_a \3) \2 \4 \Big)\\
+4 \Big( (\cd^a \1)\2(\cd_a \3)\4+ (\cd^a \1)\3(\cd_a \4) \2 -
(\cd^a \1)\3(\cd_a\2) \4 \Big) \\
+3(\cd_\a W^\b) \1 W_\b \6 +9(\cd_\a W^\b) \1 \6 W_\b +3(\cd_\a
W^\b) \6 \1 W_\b \\+3(\cd_\a W^\b) \3 \1 W_\b \4 -6(\cd_\a W^\b)
\1 \3 W_\b \4 -6(\cd_\a W^\b) \3 \1 \4 W_\b \bigg)\\+ c.c.
\end{multline}
Here the complex conjugate of any term is effectively obtained by
replacing all undotted spinor indices (and unbarred objects) by
dotted spinor indices (and barred objects) and vice-versa.
Integrating by parts, the result can be brought into the more
compact form
\begin{multline}
\mathrm{Tr}(a_5)=\frac{1}{30}\mathrm{Tr}\bigg(2 \Big( (\cd^a \cd_a
\1)\2 \6+ (\cd^a \cd_a \1)\6 \2 - (\cd^a \cd_a \1)\3 \2 \4
\Big)\\
+\Big((\cd^a \1)(\cd_a \2) \6+ (\cd^a \1)(\cd_a \3) \4 \2 - (\cd^a
\1)(\cd_a \3) \2 \4 \Big)\\ +5\Big((\cd_\a W^\b) \1 W_\b \6
-(\cd_\a W^\b) \1 \3 W_\b \4\Big) \bigg)+ c.c.
\label{eq:2ndbyparts}
\end{multline}
The corresponding piece of the one-loop effective action can
immediately be deduced by insertion into equation
(\ref{eq:effectiveaction}).

\subsection{$a_5$ at the component level \label{sec:component}}
We are now in a position to extract the component form of
$\mathrm{Tr}(a_5).$ We consider only the contribution containing
the field strength $F_{a b}$ and its covariant derivatives. It is
natural to split the result into two parts and compute their
component fields separately. Firstly, consider only terms with two
covariant derivatives:
\begin{multline}
\frac{1}{30} \mathrm{Tr}\bigg(2 \Big( (\cd^a \cd_a \1)\2 \6+
(\cd^a \cd_a \1)\6 \2 - (\cd^a \cd_a \1)\3 \2 \4 \Big)\\
+\Big((\cd^a \1)(\cd_a \2) \6+ (\cd^a \1)(\cd_a \3) \4 \2 - (\cd^a
\1)(\cd_a \3) \2 \4 \Big) \bigg)+ c.c. \no
\end{multline}
Using standard techniques (for example, see \cite{kuzenko}), it is
not difficult to show that the relevant component part of this
superfield expression is:
\begin{multline}\label{eq:ddcf}
\frac{1}{30}\mathrm{Tr} \bigg(2 \Big( (\nab^e F^{ab})(\nab_e
F_{bc})F^{cd}F_{da}+(\nab^e F^{ab})(\nab_e
F^{cd})F_{bc}F_{da}+(\nab^e F^{ab})(\nab_e F_{ca})F_{bd}F^{dc}\Big) \\
-\frac{1}{2} \Big((\nab^e F^{ab})(\nab_e
F_{ab})F^{cd}F_{cd}+(\nab^e F^{ab})(\nab_e
F^{cd})F_{ab}F_{cd}+(\nab^e F^{ab})(\nab_e F^{cd})F_{cd}F_{ab}
\Big) \\+ 4\Big( (\nabla^2 F^{ab})F_{bc}F^{cd}F_{da}+(\nab^2
F^{ab})F^{cd}F_{bc}F_{da}+(\nab^2
F^{ab})F_{ca}F_{bd}F^{dc}\Big) \\
-\Big((\nab^2 F^{ab})F_{ab}F^{cd}F_{cd}+(\nab^2 F^{ab})
F^{cd}F_{ab}F_{cd}+(\nab^2 F^{ab})F^{cd}F_{cd}F_{ab} \Big)\bigg)
\end{multline}
where
\begin{equation}
\begin{split}
\nab_a=\pd_a-\ii A_a, \qquad \qquad \qquad  [\nab_a,\nab_b]=-\ii F_{a b} \qquad \qquad \\
F_{ab}=\pd_a A_b-\pd_b A_a-\ii [A_a,A_b], \qquad
\nab_c(F_{ab})=\pd_{c}F_{ab}-\ii [A_c,F_{ab}].
\end{split}
\end{equation}
The $\nab^2 F$ terms can be converted into $F^2$-type terms using
\begin{equation}
\nab^2 F_{ab}= 2\ii(F_{ac}F^{c}_{\phantom{c}
b}-F_{bc}F^{c}_{\phantom{c} a}),
\end{equation}
thus generating some $F^5$ terms.  The resulting expression can be
further simplified by recognizing that out of the six distinct
$F^5$ structures, only four are linearly independent, as seen via
the following identities:
\begin{eqnarray}
\mathrm{Tr}(\FF)&=&\frac{1}{5}\mathrm{Tr}(\FA)-\mathrm{Tr}(\FB)
\no
\\ &&+\mathrm{Tr}(\FC)+\frac{3}{5}\mathrm{Tr}(\FD)\label{eq:F5a}
\end{eqnarray}
and
\begin{eqnarray}
\mathrm{Tr}(\FE)&=&-\frac{3}{5}\mathrm{Tr}(\FA)+\mathrm{Tr}(\FB)
\no
\\&&+\mathrm{Tr}(\FC)+\frac{1}{5}\mathrm{Tr}(\FD).\label{eq:F5b}
\end{eqnarray}

Using these, equation (\ref{eq:ddcf}) reduces to
\begin{multline}\label{eq:cf1}
\frac{1}{30}\mathrm{Tr} \bigg(2 \Big( (\nab^e F^{ab})(\nab_e
F_{bc})F^{cd}F_{da}+(\nab^e F^{ab})(\nab_e
F^{cd})F_{bc}F_{da}+(\nab^e F^{ab})(\nab_e F_{ca})F_{bd}F^{dc}\Big) \\
-\frac{1}{2} \Big((\nab^e F^{ab})(\nab_e
F_{ab})F^{cd}F_{cd}+(\nab^e F^{ab})(\nab_e
F^{cd})F_{ab}F_{cd}+(\nab^e F^{ab})(\nab_e F^{cd})F_{cd}F_{ab}
\Big) \\
+4\ii\Big(\FA+3\FB-\FC+\FD\Big)\bigg).
\end{multline}

The bosonic component of $\mathrm{Tr}(a_5)$ coming from the terms in
(\ref{eq:2ndbyparts}) with a single
covariant derivative is
\begin{multline}
-\frac{\ii}{12}\mathrm{Tr} \bigg(2\FA+4\FB-2\FC \\
+\FE+\FF \bigg),\qquad
\end{multline}
which, after the application of (\ref{eq:F5a}) and (\ref{eq:F5b})
becomes
\begin{equation}\label{eq:cf2}
\begin{array}{l}
-\frac{\ii}{15}\mathrm{Tr}\bigg(2 \FA +5 \FB+ \FD \bigg).
\end{array}
\end{equation}

Finally, adding (\ref{eq:cf1}) and (\ref{eq:cf2}), one obtains the
overall bosonic component of $\mathrm{Tr}(a_5)$,
\begin{multline}\label{eq:compresult}
\frac{1}{30}\mathrm{Tr}\bigg(2\Big((\nab^e F^{ab})(\nab_e
F_{bc})F^{cd}F_{da}+(\nab^e F^{ab})(\nab_e
F^{cd})F_{bc}F_{da}+(\nab^e F^{ab})(\nab_e F_{ca})F_{bd}F^{dc}\Big) \\
-\frac{1}{2}\Big((\nab^e F^{ab})(\nab_e
F_{ab})F^{cd}F_{cd}+(\nab^e F^{ab})(\nab_e
F^{cd})F_{ab}F_{cd}+(\nab^e F^{ab})(\nab_e F^{cd})F_{cd}F_{ab} \Big)\\
+2\ii \Big(\FB-2 \FC +\FD \Big) \bigg).
\end{multline}

After conversion to the basis used in \cite{KS1,MBM} (see Appendix
A), we find exact agreement, up to an overall multiplicative
constant, with equation (\ref{eq:NABIresult}), which describes the
$(\a')^3$ terms in the non-Abelian Born-Infeld action.

Conversely we do not find agreement with the results of
\cite{refolli}, equation (\ref{eq:refolliresult}), where the
component form of the $\mathcal{N}=4$ super Yang-Mills one-loop
effective action is extracted in several pieces through supergraph
techniques\footnote{It is perhaps worth pointing out that separate
agreement (up to  overall multiplicative constants) is found
between our equations (\ref{eq:ddcf}) and (\ref{eq:cf2}), and the
corresponding equations in \cite{refolli}, (6.2) and (4.13)
respectively. However, the multiplicative factors differ in each
case, and so on assembling the final result, a discrepancy
emerges.}.

\section{The $F^6$ terms \label{sec:a6}}
\subsection{Expanding moments}
Employing the procedure outlined above to compute $a_k$ for $k>5$
will necessarily involve asymptotically expanding moments to
higher than leading order. A prescription will therefore be
required if this scheme is to be generalized.
It is possible to appeal to a set of techniques similar to
those already seen in the previous sections.

More specifically, to evaluate any moment to arbitrary order, one
proceeds iteratively by using  the following generalizations of
the identities (\ref{eq:id1}), (\ref{eq:id2}) and (\ref{eq:id3}):
\begin{align}
0&=\int\!\! \dd \h \; \frac{\pd \phantom{k_b}}{\pd
k_b}\left(X_{A_1}\ldots X_{A_n} e^{t\D}\right) \label{eq:id1a}\\
0&=\int\!\! \dd \h \; \frac{\pd \phantom{\e_{\b}}}{\pd
\e_\b}\left(X_{A_1}\ldots X_{A_n} e^{t\D}\right) \label{eq:id2a}\\
0&=\int\!\! \dd \h \; \frac{\pd
\phantom{\ba{\e}_{\.{\b}}}}{\pd\ba{\e}_{\.{\b}}}\left(X_{A_1}\ldots
X_{A_n} e^{t\D}\right); \label{eq:id3a}
\end{align}
or by differentiation with respect to $t$ as in (\ref{eq:de1}):
\begin{equation}
\frac{d^{m} K_{A_1 \ldots A_n}(z,t)}{d t^m}= \int\!\! \dd\h
\;X_{A_1}\ldots X_{A_n}\D^m e^{t\D};
\end{equation}
or by using a combination of the two as in (\ref{eq:de2}).  Of
course, none of this actually computes the moment directly, but is
used with the intention of expressing it in terms of other moments
with the same number or more indices, which are generally easier
to compute directly. Using this procedure, expanding a moment to
some order will usually require knowledge of the expansion of
several other moments to the same or lower order. Consequently at
some point it will be necessary to evaluate at least one moment
directly by expanding the exponential.

\subsection{The moment hierarchy and $a_6$}
Computing $a_6$ involves summing from $n=1$ to 6 on the right hand
side of (\ref{eq:de3}), which generates a hierarchy of moments, a
partial list being given below (all but $K(z,t)$ required to
subleading order):
\begin{small}
\begin{gather*}
K_{\a\b\.{\a}\.{\b}a b}(z,t) \\
K_{\a\b\.{\a}\.{\b}a}(z,t) \quad K_{\a\b\.{\a}a b}(z,t) \quad
K_{\a\.{\a}\.{\b}a b}(z,t) \quad K_{\a\b\.{\a}\.{\b}}(z,t) \\
K_{\a\b\.{\a} a}(z,t) \quad K_{\a\.{\a}\.{\b}a}(z,t) \quad K_{\a\b
a b}(z,t) \quad K_{\a\.{\a}a b}(z,t) \quad K_{\.{\a}\.{\b}a
b}(z,t)
\quad K_{\a\b\.{\a}}(z,t) \quad K_{\a\.{\a}\.{\b}}(z,t)\\
K_{\a\b a}(z,t) \quad K_{\a \.{\a} a}(z,t) \quad K_{\.{\a}\.{\b}
a}(z,t) \quad K_{\a a b}(z,t) \quad K_{\.{\a} a b}(z,t) \quad
K_{\a
\b}(z,t) \quad K_{\a \.{\a}}(z,t) \quad K_{\.{\a}\.{\b}}(z,t)\\
\vdots\\
K_{a b}(z,t) \quad K_{\a a}(z,t) \quad K_{\.{\a} a}(z,t) \quad
K_{\a}(z,t) \quad K_{\.{\a}}(z,t)\\
K(z,t)\\
\end{gather*}
\end{small}
Generally speaking, the following structure is present: from top
to bottom the moments decrease in the number of indices, increase
in difficulty of expansion, and  the exponent of $t$ in the
leading order term increases (each row contains moments with the
same leading order). From left to right, the moments decrease in
their difficulty of expansion, and clearly many are related by
complex conjugation.

In the prescription outlined above, the expansion of any moment
hinges on having computed the expansion of a number of those next
to or above it in the hierarchy, so naturally one begins at the
top and works down. To be more explicit, consider the following
three examples which cover all important points.

$K_{\a\b\.{\a}\.{\b}}(z,t)$ turns out to be a rather important
object in this hierarchy, in that all others can be expressed in
terms of it. It's power series to subleading order is not
difficult to compute by directly expanding the exponential, and
takes the simple form:
\begin{equation}
K_{\a\b\.{\a}\.{\b}}(z,t)= -\frac{1}{16
\p^2}\frac{4}{t^2}\ce_{\a\b}\ce_{\.{\a}\.{\b}}+\mathcal{O}(t^0),
\end{equation}
where the $t^{-1}$ term vanishes due to the equations of motion.

Computing $K_{\a\b\.{\a}}(z,t)$ involves the identity
\begin{equation}
0=\int\!\! \dd \h \; \frac{\pd
\phantom{\ba{\e}_{\.{\g}}}}{\pd\ba{\e}_{\.{\g}}}\left(X_\a X_\b
\ba{X}_{\.{\a}}\ba{X}_{\.{\b}} e^{t\D}\right)
\end{equation}
which, after the contraction of $\.{\b}$ and $\.{\g},$ leads to
\begin{equation}\label{eq:ke1}
K_{\a\b\.{\a}}(z,t)= \int\!\!\dd \h\;X_\a X_\b
\ba{X}_{\.{\a}}\ba{X}_{\.{\b}}\sum_{n=0}^{\infty}
\frac{t^{n+1}}{(n+1)!}ad_{\D}^{\,n}(\ba{W}^{\.{\b}})\;e^{t\D}.
\end{equation}
To leading or subleading order, the summation can be truncated at
$n=0$ or 2 respectively.  Alternatively, one may have chosen to
start with the identity
\begin{equation}
0=\int\!\! \dd \h \; \frac{\pd \phantom{k_b}}{\pd k_b} \left(X_\a
X_\b \ba{X}_{\.{\a}} X_a \;e^{t\D}\right)
\end{equation}
to obtain an expression for $K_{\a\b\.{\a}}(z,t)$, but this ends
up being far more complicated. In general, if the moment in
question has less than four spinor indices, it is more convenient
to chose the identities (\ref{eq:id2a}) or (\ref{eq:id3a}) rather
than (\ref{eq:id1a}). However, if there are four spinor indices
there is no choice and (\ref{eq:id1a}) must be used.

Summing from $n=0$ to 2 in (\ref{eq:ke1}), one finds that to
subleading order, $K_{\a\b\.{\a}}(z,t)$ can be expressed in terms
of
\begin{gather*}
t^3 K_{\a\b\.{\a}\.{\b}a b}(z,t), \quad t^2
K_{\a\b\.{\a}\.{\b}}(z,t), \quad t K_{\a\b\.{\a}\.{\b}}(z,t) \quad
\textrm{and} \quad t K_{\a\b\.{\a}}(z,t),
\end{gather*}
where only $K_{\a\b\.{\a}\.{\b}}(z,t)$ is actually required to
subleading order.  Notice that $K_{\a\b\.{\a}}(z,t)$ is actually
expressed in terms of itself (multiplied by $t$).  This is a
typical feature of this approach, and one can either rely on the
fact that $K_{\a\b\.{\a}}(z,t)$ is already known to leading order,
or bring it to the left hand side and premultiply both sides by an
inverse operator (to appropriate order) to generate an new
expression for $K_{\a\b\.{\a}}(z,t)$  in terms of only
$K_{\a\b\.{\a}\.{\b}a b}(z,t)$ and $K_{\a\b\.{\a}\.{\b}}(z,t)$.

As a final example, consider expanding the moment $K_{\a\b}(z,t)$
to subleading order. In this case it is far more convenient to
differentiate with respect to $t$. The power series expansion of
$K_{\a\b}(z,t)$ will look like
\begin{equation}
K_{\a\b}(z,t)= A+ t\, B + \mathcal{O}(t^2),
\end{equation} and
so,
\begin{equation}
\frac{d K_{\a\b}(z,t)}{d t}= B + \mathcal{O}(t).
\end{equation}
Therefore, after a little work, to order unity in $t$ (ie $t^0$)
\begin{equation}
\frac{d K_{\a\b}(z,t)}{d t}= K_{\a\b a}^{\phantom{\a\b a} a}(z,t)-
\ba{W}^{\.{\a}} K_{\a\b \.{\a}}(z,t).
\end{equation}
So if both $K_{\a\b a b}(z,t)$ and $K_{\a\b \.{\a}}(z,t)$, which
are higher up the hierarchy, are known to subleading order (to
order unity in $t$), $K_{\a\b}(z,t)$ can immediately be evaluated
to subleading order (ie identify $B$). Additionally this generates
the leading order identity
\begin{equation}
K_{\a\b a}^{\phantom{\a\b a} a}(z,t)= \ba{W}^{\.{\a}} K_{\a\b
\.{\a}}(z,t),
\end{equation}
which serves as a useful consistency check.

Having summed the right hand side of (\ref{eq:de3}) from $n=1$ to
6 and expanded all surviving terms to order $t^3$, $a_6$ can be
identified (and of course $a_5$ is also recovered in this
process).  The final result is given in Appendix B. Due to its
size, and the fact that there are many equivalent ways of
presenting the result, it is a significant challenge to find the
most compact and symmetric looking expression. By extensive use of
commutation relations, equations of motion and the cyclicity of
the trace, the result is brought into a manifestly real form
involving only seven distinct types of terms, each listed
schematically below (where $G_{ab}$ was defined in
(\ref{eq:commutation})):
\begin{gather}
W^2 \; \ba{W}^2 \; \cd_a^4, \qquad W^2 \; \ba{W}^2 \; G_{ab} \;
\cd_a^2, \qquad W^3 \; \ba{W}^2 \; \cd_\a \; \cd_a^2, \qquad W^2
\; \ba{W}^3 \; \ba{\cd}_{\.{\a}} \; \cd_a^2, \qquad \no \\ W^3 \;
\ba{W}^3 \; \cd_\a \; \ba{\cd}_{\.{\a}}, \qquad W^4 \; \ba{W}^2 \;
\cd_\a^2, \qquad W^2 \; \ba{W}^4 \; \ba{\cd}_{\.{\a}}^2 \,\, . \no
\end{gather}
Here, for example, $W^2 \, \ba{W}^2 \, \cd_a^4$ is taken to mean
terms which contain (some specific permutation and contraction of)
two chiral superfield strengths, two antichiral superfield
strengths and four spacetime covariant derivatives.

Again the corresponding contribution to the effective action can
be obtained by inspection, but integrating by parts offers little
in the way of simplification. Extraction of the component form of
$a_6$ is now in principle straightforward, and contains $F^6$-type field
strength terms.

\section{Discussion \label{sec:discussion}}
In this paper, the $F^5$ and $F^6$ terms in the one-loop
low-energy effective action for $\mathcal{N}=4$ supersymmetric
Yang-Mills theory have been computed in $\mathcal{N}=1$ superfield
form. As noted in the Introduction, the $F^5$ terms are consistent
with the result (\ref{eq:NABIresult}) obtained from superstring
theory, providing evidence for a unique form for the non-Abelian
Born-Infeld action at this order. The $F^6$ terms in the
low-energy effective action have not previously been computed, and
comparison with recent superstring results \cite{koerber,ST1,ST2}
remains to be carried out.

It is possible to perform a non-trivial test on the $F^6$ results.
The form of the one-loop low-energy effective action is known in
the Abelian case in the constant field strength approximation
\cite{fradkin,buchbinder}, and the coefficient of $F^6$ is zero.
Inspection of $a_6$ reveals that in the Abelian limit, $F^6$
contributions for constant field strength can come only from terms
of the form: $W^4 \; \ba{W}^2 \; \cd_\a^2$ and $W^2 \; \ba{W}^4 \;
\ba{\cd}_{\.{\a}}^2$ (which encompass the last terms in $a_6$ as
it is given). Explicitly, the result reduces to
\begin{align}
\frac{1}{2520}\bigg(&(\cd_\a \1[\b])(\cd_\b \1)\5\6 (5+5+5
+33+19+19)\no \\& +(\cd_\a \1[\b])(\cd_\b \1)\5\6 (-6+8+8-6-6+22)
\no \\& +(\cd_\a \1[\b])(\cd_\b \1)\5\6
(9-\frac{5}{2}-\frac{5}{2}-5)\no \\& +(\cd_\a \1[\b])(\cd_\g
\1)\2[\b] \1[\g] \6(14+14+14+42+28+28) \no
\\& +14(\cd_\a\1[\b])(\cd_\g \1)\2[\b] \1[\g] \6(1+1+2)
+14(\cd_\a\1[\b])(\cd_\g \1)\2[\b] \1[\g] \6\bigg)+c.c.
\no\\=\frac{1}{24}\bigg(&(\cd_\a \1[\b])(\cd_\b
\1)\5\6+2(\cd_\a\1[\b])(\cd_\g \1)\2[\b] \1[\g] \6\bigg)+c.c.\no
\end{align}
The second term can be rearranged into the same form
as the first, since in the Abelian case,
\begin{eqnarray}
\2 \2[\b]&=&\frac{1}{2}\ce_{\a\b}\5 \no\\
\Rightarrow(\cd_\a\1[\b])(\cd_\g \1)\2[\b] \1[\g]
\6&=&-\frac{1}{2}(\cd_\a\1[\b])(\cd_\b \1)\5\6,\no
\end{eqnarray}
and indeed one finds non-trivial cancellation, consistent with
\cite{fradkin,buchbinder}.

\acknowledgments

I wish to thank I.N. McArthur for discussions, ideas, suggestions
and explanations.  I am also grateful to S.M. Kuzenko for
suggestions and references.

\appendix

\section{Change of basis} In this appendix we briefly outline the
transformation from the basis used in our result
(\ref{eq:compresult}), to that of (\ref{eq:NABIresult}).  For
simplicity we introduce the following notation:
\begin{align*}
s_{0,0}&=\mathrm{Tr}(\FA) &
s_{0,1}&=\mathrm{Tr}(\FB) \\
s_{0,2}&=\mathrm{Tr}(\FC) &
s_{0,3}&=\mathrm{Tr}(\FD) \\
s_{0,4}&=\mathrm{Tr}(\FE) &
s_{0,5}&=\mathrm{Tr}(\FF) \\
s_{1,0}&=\mathrm{Tr}(\FGa) &
s_{1,1}&=\mathrm{Tr}(\FGb) \\
s_{1,2}&=\mathrm{Tr}(\FGc) &
s_{1,3}&=\mathrm{Tr}(\FMa) \\
s_{1,4}&=\mathrm{Tr}(\FMb) &
s_{1,5}&=\mathrm{Tr}(\FMc) \\
s_{1,6}&=\mathrm{Tr}(\FMd) &
s_{1,7}&=\mathrm{Tr}(\FGd) \\
s_{2,0}&=\mathrm{Tr}(\FLa) &
s_{2,1}&=\mathrm{Tr}(\FLc) \\
s_{2,2}&=\mathrm{Tr}(\FLb).
\end{align*}

Using the equations of motion, the Bianchi identity, integration
by parts and the cyclic property of the trace, one can establish
the following identities:
\begin{align*}
s_{0,4}=& -\frac{3}{5}s_{0,0}+s_{0,1}+s_{0,2}+\frac{1}{5}s_{0,3}&
s_{0,5}=& \frac{1}{5}s_{0,0}-s_{0,1}+s_{0,2}+\frac{3}{5}s_{0,3}\\
s_{1,1}=&-\frac{1}{2} s_{1,7}+ 2\ii s_{0,5}&
s_{1,2}=&-s_{1,0}-4s_{2,2}+4\ii s_{0,4} \\
s_{1,3}=&-\ii s_{0,0}-\ii s_{0,1}+\ii s_{0,2}+3\ii s_{0,3} &
s_{1,5}=&-s_{1,6}-s_{1,4}-2\ii s_{0,1}+2\ii s_{0,2}\\
&-\frac{\ii}{2}s_{0,4}-\ii s_{0,5}-2s_{1,4}-s_{1,6}
\\&+\frac{3}{8}s_{1,7}-4s_{2,0}+4s_{2,1}-s_{2,2}
\end{align*}

In this notation, the bosonic component of $\mathrm{Tr}(a_5)$,
equation (\ref{eq:compresult}), takes the form:
\begin{equation}
\frac{1}{30}\mathrm{Tr}(2(s_{1,3}+s_{1,4}+s_{1,5})-\frac{1}{2}(s_{1,0}
+s_{1,1}+s_{1,2})+2\ii(s_{0,1}-2s_{0,2}+s_{0,3})).
\end{equation}
Using the above relations, elimination of
$s_{1,1},\,s_{1,2},\,s_{1,3}$ and $s_{1,5}$, followed by the
further elimination of $s_{0,4}$ and $s_{0,5}$, yields the
following expression:
\begin{equation}
\frac{4}{15}\mathrm{Tr}(s_{2,1}-s_{2,0}-\frac{1}{2}s_{1,4}
-\frac{1}{2}s_{1,6}+\frac{1}{8}s_{1,7}
-\frac{\ii}{10}s_{0,0}-\frac{\ii}{2}s_{0,1}-\frac{\ii}{2}s_{0,2}
+\frac{7\ii}{10}s_{0,3}).
\end{equation}
which up to an overall multiplicative factor, is equation
(\ref{eq:NABIresult}).

\newpage
\section{$\mathrm{Tr}(a_6)$}

\begin{small}
\begin{eqnarray}\label{eq:result} \no 
\begin{array}{l}
\mathrm{Tr}( a_6)= \\ \frac{1}{2520}\mathrm{Tr} \bigg(28
\perm{\cda{\cd^a\cd_a\cd^b\cd_b}}{}{}{} \\ +
62\perm{\cda{\cd^a\cd^b\cd_b}}{\cda{\cd_a}}{}{}\\ +
62\perm{\cda{\cd^a\cd^b\cd_b}}{}{}{\cda{\cd_a}}\\ +
44\perm{\cda{\cd^a\cd^b\cd_b}}{}{\cda{\cd_a}}{}\\ +
48\perm{\cda{\cd^b\cd_b}}{\cda{\cd^a\cd_a}}{}{}\\ +
22\perm{\cda{\cd^b\cd_b}}{}{\cda{\cd^a\cd_a}}{}\\ +
50\perm{\cda{\cd^b\cd_b}}{\cda{\cd^a}}{\cda{\cd_a}}{}\\ +
50\perm{\cda{\cd^b\cd_b}}{}{\cda{\cd^a}}{\cda{\cd_a}}\\ +
40\perm{\cda{\cd^b\cd_b}}{\cda{\cd^a}}{}{\cda{\cd_a}}\\ +
44\perm{\cda{\cd^a\cd^b}}{\cda{\cd_a\cd_b}}{}{}\\ +
12\perm{\cda{\cd^a\cd^b}}{}{\cda{\cd_a\cd_b}}{}\\ +
52\perm{\cda{\cd^a\cd^b}}{\cda{\cd_a}}{\cda{\cd_b}}{}\\ +
52\perm{\cda{\cd^a\cd^b}}{}{\cda{\cd_a}}{\cda{\cd_b}}\\ +
64\perm{\cda{\cd^a\cd^b}}{\cda{\cd_a}}{}{\cda{\cd_b}}\\ +
9\perm{\cda{\cd^a}}{\cda{\cd^b}}{\cda{\cd_a}}{\cda{\cd_b}} \\ +
26\perm{\cda{\cd^a}}{\cda{\cd_a}}{\cda{\cd^b}}{\cda{\cd_b}}\\


+ 84\, G^{ab}\perm{\cda{\cd_a}}{\cda{\cd_b}}{}{}\\
+ 84\, G^{ab}\perm{\cda{\cd_a}}{}{}{\cda{\cd_b}}\\
+ 60\, G^{ab}\perm{\cda{\cd_a}}{}{\cda{\cd_b}}{}\\
- 4\, G^{ab}\perm{}{\cda{\cd_a}}{\cda{\cd_b}}{}\\
- 4\, G^{ab}\perm{}{\cda{\cd_a}}{}{\cda{\cd_b}}\\
+ 32\, G^{ab}\perm{}{}{\cda{\cd_a}}{\cda{\cd_b}}\\


+(\cd^a\cd_a\cd_\a \1[\b]) \Big( 16  \1  \2[\b]  \6 +18 \3  \1
\2[\b]  \4 +16 \6          \1      \2[\b] \\
\qquad\qquad\qquad\qquad +50 \1  \6              \2[\b] -34 \1  \3
\2[\b]  \4
-34 \3  \1      \4      \2[\b]\Big) \\


+(\cd^a\cd_\a \1[\b]) \Big( 16  (\cd_a\1)  \2[\b]  \6 +24
(\cd_a\3)  \1      \2[\b]  \4 +24 (\cd_a\3)  \4      \1
\2[\b] \\\qquad\qquad\qquad\qquad +64 (\cd_a\1)  \6
\2[\b] -40 (\cd_a\1)  \3      \2[\b]  \4
-48 (\cd_a\3)  \1      \4      \2[\b]\Big) \\

+(\cd^a\cd_\a \1[\b]) \Big( 8   \1  (\cd_a\2[\b])  \6 +12 \3
(\cd_a\1)      \2[\b]  \4 +16 \3  (\cd_a\4)      \1      \2[\b]
\\\qquad\qquad\qquad\qquad +36 \1  (\cd_a\3)      \4      \2[\b]
-20 \1  (\cd_a\3)      \2[\b]  \4
-28 \3  (\cd_a\1)      \4      \2[\b]\Big)\\

+(\cd^a\cd_\a \1[\b]) \Big( 16  \1  \2[\b]  (\cd_a\3)      \4 +12
\3  \1      (\cd_a\2[\b])  \4 +8  \6          (\cd_a\1)
\2[\b] \\\qquad\qquad\qquad\qquad +36 \1  \3      (\cd_a\4)
\2[\b] -28 \1  \3      (\cd_a\2[\b])  \4
-20 \3  \1      (\cd_a\4)      \2[\b]\Big)\\

+(\cd^a\cd_\a \1[\b]) \Big( 24  \1  \2[\b]  \3      (\cd_a\4) +24
\3  \1      \2[\b]  (\cd_a\4) +16 \6          \1
(\cd_a\2[\b]) \\\qquad\qquad\qquad\qquad +64 \1  \6
(\cd_a\2[\b]) -48 \1  \3      \2[\b]  (\cd_a\4)
-40 \3  \1      \4      (\cd_a\2[\b])\Big)\\


+(\cd_\a \1[\b]) \Big( 10  (\cd^a\cd_a\1)  \2[\b]  \6 +16
(\cd^a\cd_a\3)  \1      \2[\b]  \4 +18 (\cd^a\cd_a\3)  \4      \1
\2[\b] \\\qquad\qquad\qquad\qquad +44 (\cd^a\cd_a\1)  \6
\2[\b] -26 (\cd^a\cd_a\1)  \3      \2[\b]  \4 -34 (\cd^a\cd_a\3)
\1      \4      \2[\b]\Big)\no
\end{array}
\end{eqnarray}

\newpage
\begin{eqnarray}
\begin{array}{l}
\no 
+(\cd_\a \1[\b]) \Big( 10  \1  (\cd^a\cd_a\2[\b])  \6 +10 \3
(\cd^a\cd_a\1)      \2[\b]  \4 +16 \3  (\cd^a\cd_a\4)      \1
\2[\b] \\\qquad\qquad\qquad\qquad +36 \1  (\cd^a\cd_a\3)      \4
\2[\b] -20 \1  (\cd^a\cd_a\3)      \2[\b]  \4
-26 \3  (\cd^a\cd_a\1)      \4      \2[\b]\Big) \\

+(\cd_\a \1[\b]) \Big( 16  \1  \2[\b]  (\cd^a\cd_a\3)      \4 +10
\3  \1      (\cd^a\cd_a\2[\b])  \4 +10 \6          (\cd^a\cd_a\1)
\2[\b] \\ \qquad\qquad\qquad\qquad +36 \1  \3      (\cd^a\cd_a\4)
\2[\b] -26 \1  \3      (\cd^a\cd_a\2[\b])  \4
-20 \3  \1      (\cd^a\cd_a\4)      \2[\b]\Big) \\

+(\cd_\a \1[\b]) \Big( 18  \1  \2[\b]  \3      (\cd^a\cd_a\4) +16
\3  \1      \2[\b]  (\cd^a\cd_a\4) +10 \6          \1
(\cd^a\cd_a\2[\b]) \\ \qquad\qquad\qquad\qquad +44 \1  \6
(\cd^a\cd_a\2[\b]) -34 \1  \3      \2[\b]  (\cd^a\cd_a\4)
-26 \3  \1      \4      (\cd^a\cd_a\2[\b])\Big) \\


+(\cd_\a \1[\b]) \Big( 4   (\cd^a\1)  (\cd_a\2[\b])  \6 +16
(\cd^a\3)  (\cd_a\1)      \2[\b]  \4 +24 (\cd^a\3)  (\cd_a\4)
\1      \2[\b] \\ \qquad\qquad\qquad\qquad +44 (\cd^a\1)
(\cd_a\3)      \4      \2[\b] -20 (\cd^a\1)  (\cd_a\3)      \2[\b]
\4
-40 (\cd^a\3)  (\cd_a\1)      \4      \2[\b]\Big) \\

+(\cd_\a \1[\b]) \Big( 8   (\cd^a\1)  \2[\b]  (\cd_a\3)      \4 +8
(\cd^a\3)  \1      (\cd_a\2[\b])  \4 +12 (\cd^a\3)  \4
(\cd_a\1)      \2[\b] \\ \qquad\qquad\qquad\qquad +28 (\cd^a\1)
\3      (\cd_a\4)      \2[\b] -16 (\cd^a\1)  \3      (\cd_a\2[\b])
\4
-20 (\cd^a\3)  \1      (\cd_a\4)      \2[\b]\Big) \\

+(\cd_\a \1[\b]) \Big( 12  (\cd^a\1)  \2[\b]  \3      (\cd_a\4)
+16 (\cd^a\3)  \1      \2[\b]  (\cd_a\4) +12 (\cd^a\3)  \4      \1
(\cd_a\2[\b]) \\ \qquad\qquad\qquad\qquad +40 (\cd^a\1)  \6
(\cd_a\2[\b]) -28 (\cd^a\1)  \3      \2[\b]  (\cd_a\4)
-28 (\cd^a\3)  \1      \4      (\cd_a\2[\b])\Big) \\

+(\cd_\a \1[\b]) \Big( 16  \1  (\cd^a\2[\b])  (\cd_a\3)      \4 +4
\3  (\cd^a\1)      (\cd_a\2[\b])  \4 +16 \3  (\cd^a\4)
(\cd_a\1)      \2[\b] \\ \qquad\qquad\qquad\qquad +36 \1
(\cd^a\3)      (\cd_a\4)      \2[\b] -20 \1  (\cd^a\3)
(\cd_a\2[\b])  \4
-20 \3  (\cd^a\1)      (\cd_a\4)      \2[\b]\Big) \\

+(\cd_\a \1[\b]) \Big( 12  \1  (\cd^a\2[\b])  \3      (\cd_a\4) +8
\3  (\cd^a\1)      \2[\b]  (\cd_a\4) +8  \3  (\cd^a\4)      \1
(\cd_a\2[\b]) \\ \qquad\qquad\qquad\qquad +28 \1  (\cd^a\3)
\4       (\cd_a\2[\b]) -20 \1  (\cd^a\3)      \2[\b]  (\cd_a\4)
-16 \3  (\cd^a\1)      \4      (\cd_a\2[\b])\Big) \\

+(\cd_\a \1[\b]) \Big( 24  \1  \2[\b]  (\cd^a\3)      (\cd_a\4)
+16 \3  \1      (\cd^a\2[\b])  (\cd_a\4) +4  \6          (\cd^a\1)
(\cd_a\2[\b]) \\ \qquad\qquad\qquad\qquad +44 \1  \3
(\cd^a\4)      (\cd_a\2[\b]) -40 \1  \3      (\cd^a\2[\b])
(\cd_a\4)
-20 \3  \1      (\cd^a\4)      (\cd_a\2[\b])\Big)\\


+14 (\cd_\a \1[\b])(\ba{\cd}_{\.{\a}}\3[\b])\Big(
      \1  \4      \4[\b]          \2[\b]
+   \4  \1      \2[\b]          \4[\b] -   \1  \4      \2[\b]
\4[\b]
-   \4  \1      \4[\b]          \2[\b]\Big)\\

+14 (\cd_\a \1[\b])\Big(
      \1  (\ba{\cd}_{\.{\a}}\3[\b]) \4      \4[\b]    \2[\b]
+   \4  (\ba{\cd}_{\.{\a}}\3[\b]) \1      \4[\b]
\2[\b]\\\qquad\qquad\qquad\qquad\qquad\qquad\qquad\qquad\qquad\qquad
-   \1  (\ba{\cd}_{\.{\a}}\3[\b]) \4      \2[\b]    \4[\b]
-   \4  (\ba{\cd}_{\.{\a}}\3[\b]) \4[\b]  \1        \2[\b]\Big)\\

+14 (\cd_\a \1[\b])\Big(
      \4   \1  (\ba{\cd}_{\.{\a}}\3[\b]) \4[\b]  \2[\b]
-   \1   \4  (\ba{\cd}_{\.{\a}}\3[\b]) \4[\b]  \2[\b]\Big)\\


+7(\ba{\cd}_{\.{\a}} \cd_\a \1[\b])\1 \Big( 2 \3[\b]    \4  \2[\b]
\4[\b] + \3[\b]    \4  \4[\b]  \2[\b] + \2[\b] \4 \6 \\
\qquad\qquad\qquad\qquad\qquad\qquad\qquad\qquad\qquad\qquad - 2
\4 \3[\b] \2[\b] \4[\b] - 2 \4 \6            \2[\b]
-   \3[\b] \2[\b] \3 \4[\b]\Big) \\

+7(\ba{\cd}_{\.{\a}} \cd_\a \1[\b])\4 \Big( 6   \1  \3[\b]  \2[\b]
\4[\b] +   \6  \1      \2[\b] - 3 \1  \2[\b]  \6 - 3 \3[\b]
\1      \2[\b]   \4[\b]
- 2 \1  \6      \2[\b]  \Big)\\

+7 (\ba{\cd}_{\.{\a}} \cd_\a \1[\b])\3[\b] \Big( 2   \4  \1
\2[\b]  \4[\b] +   \1  \2[\b]  \4      \4[\b] +   \4  \1
\4[\b]  \2[\b] -3  \1  \4      \2[\b]  \4[\b]
-   \4  \4[\b]  \1      \2[\b]\Big)\\


+(\cd_\a \1[\b])(\cd_\b \1) \Big( 5  \5 \6 +5 \3 \5 \4 +5 \6 \5 \\
\qquad\qquad\qquad\qquad\qquad\qquad\qquad\qquad +33 \1[\g]  \6
\2[\g] -19 \1[\g]  \3  \2[\g] \4
-19 \3 \1[\g] \4  \2[\g]\Big) \\

+(\cd_\a \1[\b])\Big( 6  \1[\g]   (\cd_\b \1)     \3       \2[\g]
\4 +8  \3       (\cd_\b \1)     \5       \4 +8  \3       (\cd_\b
\1)     \4       \5    \\ \qquad\qquad\qquad\qquad -6   \1[\g]
(\cd_\b \1)     \2[\g]   \6 -6  \1[\g]   (\cd_\b \1)     \6
\2[\g]
-22 \3       (\cd_\b \1)     \1[\g]   \4        \2[\g]\Big) \\

+(\cd_\a \1[\b])\Big( 9   \5       (\cd_\b \1)     \6 +\frac{5}{2}
\1[\g] \3      (\cd_\b \1) \2[\g] \4   \\
\qquad\qquad\qquad\qquad\qquad\qquad\qquad\qquad +\frac{5}{2} \3
\1[\g]  (\cd_\b \1) \4     \2[\g] -5  \1[\g]   \3 (\cd_\b \1)
\4       \2[\g]
\Big) \\

+(\cd_\a \1[\b])(\cd_\g \1) \Big( 14  \2[\b]  \1[\g] \6 +14 \3
\2[\b] \1[\g]   \4 +14 \6             \2[\b]   \1[\g]    \\
\qquad\qquad\qquad\qquad\qquad\qquad\qquad\qquad +42 \2[\b]  \6
\1[\g] -28 \2[\b]  \3     \1[\g]   \4
-28 \3      \2[\b] \4       \1[\g]\Big) \\

+14(\cd_\a \1[\b])\3(\cd_\g \1)\Big( \2[\b] \1[\g] \4 + \4 \2[\b]
\1[\g] -2 \2[\b] \4 \1[\g] \Big)

+14(\cd_\a \1[\b])\6(\cd_\g \1) \2[\b] \1[\g]\bigg)

+c.c. \no
\end{array}
\end{eqnarray}
\end{small}


\begin{thebibliography}{99}

\bibitem{BRS}
E. Bergshoeff, M. Rakowski and E. Sezgin, Phys. Lett.
\textbf{B185} (1987) 371.

\bibitem{MR}
R.R. Metsaev and M.A. Rakhmanov, Phys. Lett. \textbf{B193} (1987)
202.

\bibitem{MRT}
R.R. Metsaev, M.A. Rakhmanov and A.A. Tseytlin,  Phys. Lett.
\textbf{B193} (1987) 207.

\bibitem{T0} A.A. Tseytlin, Nucl. Phys. Proc. Suppl. \textbf{68} (1998)
99,
[\href{http://xxx.lanl.gov/abs/hep-th?9709123}{hep-th/9709123}].

\bibitem{PSS1}
S. Paban, S. Sethi and M. Stern, JHEP \textbf{9806} (1998) 012,
[\href{http://xxx.lanl.gov/abs/hep-th?9806028}{hep-th/9806028}].

\bibitem{CNT1}
M. Cederwall, B.E.W. Nilsson and D. Tsimpis, JHEP \textbf{0106}
(2001) 034,
[\href{http://xxx.lanl.gov/abs/hep-th?0102009}{hep-th/0102009}].

\bibitem{T1}
A.A. Tseytlin, \textit{Born-Infeld action, supersymmetry and
string theory,} in the Yuri Golfand memorial volume: The many
faces of the superworld, ed. M. Shifman, World Scientific, 2000,
[\href{http://xxx.lanl.gov/abs/hep-th?9908105}{hep-th/9908105}].

\bibitem{FT}
E.S. Fradkin and A.A. Tseytlin, Phys. Lett. \textbf{B163} (1985)
123.

\bibitem{L}
R.G. Leigh, Mod. Phys. Lett. \textbf{A4} (1989) 2767.

\bibitem{T2}
A.A. Tseytlin, Nucl. Phys. \textbf{B501} (1997) 41,
[\href{http://xxx.lanl.gov/abs/hep-th?9701125}{hep-th/9701125}].

\bibitem{W1}
E. Witten, Nucl. Phys. \textbf{B460} (1995) 435,
[\href{http://xxx.lanl.gov/abs/hep-th?9510135}{hep-th/9510135}].

\bibitem{refolli}
A. Refolli, A. Santambrogio, N. Terzi and D. Zanon, Nucl. Phys.
\textbf{B613} (2001) 64,
[\href{http://xxx.lanl.gov/abs/hep-th?0105277}{hep-th/0105277}];
Fortsch. Phys. \textbf{50} (2002) 952,
[\href{http://xxx.lanl.gov/abs/hep-th?0201106}{hep-th/0201106}].

\bibitem{KS1}
P. Koerber and A. Sevrin, JHEP \textbf{0110} (2001) 003,
[\href{http://xxx.lanl.gov/abs/hep-th?0108169}{hep-th/0108169}].

\bibitem{MBM}
R. Medina, F.T. Brandt and F.R. Machado, JHEP \textbf{0207} (2002)
071,
[\href{http://xxx.lanl.gov/abs/hep-th?0208121}{hep-th/0208121}].

\bibitem{FKS}
L. De Fosse, P. Koerber and A. Sevrin, Nucl. Phys. \textbf{B603}
(2001) 413,
[\href{http://xxx.lanl.gov/abs/hep-th?0103015}{hep-th/0103015}].

\bibitem{tseytlin}
R.R. Metsaev and A.A. Tseytlin, Nucl. Phys. \textbf{B298} (1988)
109.

\bibitem{B}
A. Bilal, Nucl. Phys. \textbf{B618} (2001) 21,
[\href{http://xxx.lanl.gov/abs/hep-th?0106062}{hep-th/0106062}].

\bibitem{K}
Y. Kitazawa, Nucl. Phys. \textbf{B289} (1987) 599.

\bibitem{vdv}
A.E. van de Ven, Nucl. Phys. \textbf{B250} (1985) 593.

\bibitem{KS2}
P. Koerber and A. Sevrin, JHEP \textbf{0109} (2001) 009,
[\href{http://xxx.lanl.gov/abs/hep-th?0109030}{hep-th/0109030}].

\bibitem{REKS}
M. de Roo, M.G.C. Eenink, P. Koerber and A. Sevrin, JHEP
\textbf{0208} (2002) 011,
[\href{http://xxx.lanl.gov/abs/hep-th?0207015}{hep-th/0207015}].

\bibitem{GW}
D.J. Gross and E. Witten, Nucl. Phys. \textbf{B277} (1986) 1.

\bibitem{cederwall}
M. Cederwall, B.E.W. Nilsson and D. Tsimpis, JHEP \textbf{0107}
(2001) 042,
[\href{http://xxx.lanl.gov/abs/hep-th?0104236}{hep-th/0104236}];
JHEP \textbf{0202} (2002) 009,
[\href{http://xxx.lanl.gov/abs/hep-th?0110069}{hep-th/0110069}];
\textit{Spinorial cohomology of Abelian D=10 super-Yang-Mills at
$\a'^3$},
[\href{http://xxx.lanl.gov/abs/hep-th?0205165}{hep-th/0205165}].

\bibitem{deroo}
A. Collinucci, M. De Roo and M.G.C. Eenink, JHEP \textbf{0206}
(2002) 024,
[\href{http://xxx.lanl.gov/abs/hep-th?0205150}{hep-th/0205150}].

\bibitem{STT}
A. Sevrin, J. Troost and W. Troost,  Nucl. Phys. \textbf{B603}
(2001) 389,
[\href{http://xxx.lanl.gov/abs/hep-th?0101192}{hep-th/0101192}].

\bibitem{mcarthur}
I.N. McArthur and T.D. Gargett, Nucl. Phys. \textbf{B497} (1997)
525,
[\href{http://xxx.lanl.gov/abs/hep-th?9705200}{hep-th/9705200}].

\bibitem{gargett}
T.D. Gargett and I.N. McArthur, J. Math. Phys. \textbf{39} (1998)
4430.

\bibitem{pletnev}
N.G. Pletnev and A.T. Banin, Phys. Rev.  \textbf{D60} (1999)
105017,
[\href{http://xxx.lanl.gov/abs/hep-th?9811031}{hep-th/9811031}].

\bibitem{banin}
A.T. Banin, I.L. Buchbinder and N.G. Pletnev, Nucl. Phys.
\textbf{B598} (2001) 371,
[\href{http://xxx.lanl.gov/abs/hep-th?0008167}{hep-th/0008167}];
Phys. Rev. \textbf{D66} (2002) 045021,
[\href{http://xxx.lanl.gov/abs/hep-th?0205034}{hep-th/0205034}].

\bibitem{koerber}
P. Koerber and A. Sevrin, \textit{The non-abelian D-brane
effective action through order $\a'{}^4$},
[\href{http://xxx.lanl.gov/abs/hep-th?0208044}{hep-th/0208044}].

\bibitem{ST1}
S. Stieberger and T.R. Taylor, \textit{Non-Abelian Born-Infeld
action and type I - heterotic duality.  I: Heterotic $F^6$ terms
at two loops},
[\href{http://xxx.lanl.gov/abs/hep-th?0207026}{hep-th/0207026}].

\bibitem{ST2}
S. Stieberger and T.R. Taylor, \textit{Non-Abelian Born-Infeld
action and type I - heterotic duality.  II: Nonrenormalization
theorems},
[\href{http://xxx.lanl.gov/abs/hep-th?0209064}{hep-th/0209064}].

\bibitem{BPT}
I.L. Buchbinder, A.Y. Petrov and A.A. Tseytlin, Nucl. Phys.
\textbf{B621} (2002) 179
[\href{http://xxx.lanl.gov/abs/hep-th?0110173}{hep-th/0110173}].

\bibitem{Wess}
J. Wess and J. Bagger, \textit{Supersymmetry and Supergravity}
(Princeton University Press, Princeton, 1991, 2nd Edition).

\bibitem{kuzenko}
I.L. Buchbinder and S.M. Kuzenko, \textit{Ideas and Methods of
Supersymmetry and Supergravity} (IOP Publishing Ltd., Bristol,
1998, 2nd Edition).

\bibitem{grisaru}
M.T. Grisaru, W. Siegel and M. Rocek, Nucl. Phys. \textbf{B159}
(1979) 429.

\bibitem{grisaru2}
M.T. Grisaru and W. Siegel, Nucl. Phys. \textbf{B187} (1981) 149.

\bibitem{GGRS}
S.J. Gates Jr, M.T. Grisaru, M. Rocek and W. Siegel,
\textit{Superspace} (Benjamin-Cummings, Reading MA, 1983),
[\href{http://xxx.lanl.gov/abs/hep-th?0108200}{hep-th/0108200}].

\bibitem{ohrndorf}
T. Ohrndorf, Phys. Lett. \textbf{B176} (1986) 421.

\bibitem{fradkin}
E.S. Fradkin and A.A. Tseytlin, Nucl. Phys. \textbf{B277} (1983)
252; Phys. Lett. \textbf{B123} (1983) 231.

\bibitem{buchbinder}
I.L. Buchbinder, S.M. Kuzenko and A.A. Tseytlin, Phys. Rev.
\textbf{D62} (2000) 045001,
[\href{http://xxx.lanl.gov/abs/hep-th?9911221}{hep-th/9911221}].

\end{thebibliography}
\end{document}